\documentclass[aps,prb,twocolumn,amsmath,amssymb,superscriptaddress]{revtex4-2}
 
\usepackage{graphicx}
\usepackage{dcolumn}
\usepackage{bm}
\usepackage[colorlinks=true, linkcolor=blue, citecolor=blue, urlcolor=blue]{hyperref}
\usepackage{xcolor}
\usepackage{comment}
\usepackage{lineno}
\begin{document}

\title{Dissecting superconductivity in the Ruddlesden-Popper nickelates: The role of electron correlation and interlayer magnetic exchange}

\author{Xiaoyang Chen}
\thanks{Equal contribution}
\affiliation{Diamond Light Source, Harwell Campus, Didcot, OX11 0DE, United Kingdom}

\author{Zezhong Li}
\thanks{Equal contribution}
\email{zzli@ustc.edu.cn}
\affiliation{National Synchrotron Radiation Laboratory and School of Nuclear Science and Technology, University of Science and Technology of China, Hefei 230026, China} 

\author{Mei Xie}
\affiliation{National Synchrotron Radiation Laboratory and School of Nuclear Science and Technology, University of Science and Technology of China, Hefei 230026, China} 

\author{Deyuan Hu}
\affiliation{Guangdong Provincial Key Laboratory of Magnetoelectric Physics and Devices,  School of Physics, Sun Yat-Sen University, Guangzhou 510275, China.}

\author{Yiu-Fung Chiu}
\affiliation{Diamond Light Source, Harwell Campus, Didcot, OX11 0DE, United Kingdom}
\affiliation{Department of Physics, University of Oxford, Clarendon Laboratory, Parks Road, Oxford OX1 3PU, United Kingdom}

\author{Stefano Agrestini}
\affiliation{Diamond Light Source, Harwell Campus, Didcot, OX11 0DE, United Kingdom}

\author{Wenliang Zhang}
\affiliation{National Synchrotron Radiation Laboratory and School of Nuclear Science and Technology, University of Science and Technology of China, Hefei 230026, China} 

\author{Yi Lu}
\affiliation{National Laboratory of Solid State Microstructures and Department of Physics, Nanjing University, Nanjing 210093, China}
\affiliation{Collaborative Innovation Center of Advanced Microstructures, Nanjing University, Nanjing 210093, China}

\author{Meng Wang}
\email{wangmeng5@mail.sysu.edu.cn}
\affiliation{Guangdong Provincial Key Laboratory of Magnetoelectric Physics and Devices,  School of Physics, Sun Yat-Sen University, Guangzhou 510275, China.}

\author{Mirian Garcia-Fernandez}
\affiliation{Diamond Light Source, Harwell Campus, Didcot, OX11 0DE, United Kingdom}

\author{Donglai Feng}
\email{dlfeng@ustc.edu.cn}
\affiliation{New Cornerstone Laboratory, Hefei National Laboratory, Hefei, 230088, China. }

\author{Ke-Jin~Zhou}
\email{kjzhou@ustc.edu.cn}
\affiliation{National Synchrotron Radiation Laboratory and School of Nuclear Science and Technology, University of Science and Technology of China, Hefei 230026, China}

\date{\today}

\begin{abstract}
	
	The discovery of superconductivity in the Ruddlesden-Popper (RP) nickelates has opened a new chapter in the search for high superconducting transition temperatures ($T_\mathrm{c}$) materials. A central and puzzling feature of this family is the wide variation in $T_\mathrm{c}$ despite their common NiO$_2$ building blocks, as highlighted by the recent observation of superconductivity at $\sim$~30 K in trilayer $\mathrm{La_4Ni_3O_{10}}$ —significantly lower than 80 K reported in bilayer $\mathrm{La_3Ni_2O_7}$. Understanding the factors that control $T_\mathrm{c}$ in this family is therefore of paramount importance. Here, we use resonant inelastic x-ray scattering (RIXS) to investigate the electronic and magnetic excitations of $\mathrm{La_4Ni_3O_{10}}$ in direct comparison with its bilayer counterpart. Our results reveal a markedly different landscape. $\mathrm{La_4Ni_3O_{10}}$ exhibits a more itinerant character, evidenced by broader Ni $dd$ orbital excitations and a strong Ni 3$d$ fluorescence continuum, suggesting weaker electronic correlations than in the bilayer. Despite this, well-defined collective spin excitations persist, including dispersive acoustic and optical magnon branches alongside an incommensurate spin density wave. Using linear spin wave theory, we extract the interlayer superexchange interaction ($J_z$) to be $\sim$~22 meV, much smaller than that in $\mathrm{La_3Ni_2O_7}$. The weaker correlation and reduced interlayer exchange together provide a consistent explanation for the substantially lower $T_\mathrm{c}$ in the trilayer compound. Our findings establish interlayer magnetic coupling and electronic correlation as key parameters governing superconductivity in layered nickelates and offer critical constraints for understanding the pairing mechanism in this emerging family.
	
\end{abstract}

\maketitle

\newpage
Since the discovery of superconductivity at 80 K in $\mathrm{La_3Ni_2O_7}$, the Ruddlesden-Popper (RP) phase nickelates (La,Pr)$_{n+1}$Ni$_n$O$_{3n+1}$ have emerged as a new platform to explore unconventional superconductivity~\cite{hsun2023,nwang2024,gwang2024,yzhu2024,mzhang2025b,qli2024,xhuang2024,mshi2025,chuang2025,ko2025signatures,zhou2025ambient}. These materials are all constructed from NiO$_2$-based multilayer building blocks and host multiorbital low-energy electronic structures involving Ni $3d_{x^2-y^2}$ and $3d_{z^2}$ orbitals with orbital-dependent electronic correlations. Moreover, at ambient pressure, a spin/charge density-wave is widely observed in RP nickelates. Despite these common features, their superconducting transition temperatures under high pressure differ dramatically, reaching $\sim$~80~K in bilayer $\mathrm{La_3Ni_2O_7}$ but only $\sim$~30~K in trilayer $\mathrm{La_4Ni_3O_{10}}$. Theoretical and experimental studies have suggested that electron–phonon coupling (EPC) appears insufficient to account for the difference in transition temperatures~\cite{yli2025,xdu2025}. The absence of superconductivity in the tetragonal phase $\mathrm{La_3Ni_2O_7}$ and $\mathrm{La_4Ni_3O_{10}}$ without SDW order at ambient pressure further underscores the central role of magnetism in enabling superconductivity in this family~\cite{mshi2025a,mshi2025b}, consistent with an unconventional pairing scenario. Therefore, determining what distinguishes the bilayer and trilayer nickelates is a central problem in understanding superconductivity in the layered nickelates.

Theoretically, $\mathrm{La_3Ni_2O_7}$ has been proposed to host a sizable interlayer antiferromagnetic exchange interaction $J_z$ originating from the bonding between Ni $3d_{z^2}$ orbitals mediated by the apical O in the NiO$_2$ bilayer~\cite{clu2024,ycao2024,xqu2024}. Under pressure, the Ni-O-Ni bonding angle along the $c$ axis changes from 168$^\circ$ to 180$^\circ$ through a structure transition, which is expected to greatly enhance the interlayer exchange~\cite{hsun2023}. Within this framework, an interlayer-coupling-driven Cooper pairing has been suggested to account for the high temperature superconductivity~\cite{lechermann2023,qyang2023,yliu2023,oh2023,clu2024,ycao2024,xqu2024,sakakibara2024a,wwu2024}. Experimentally, Ni $L_3$-edge RIXS on $\mathrm{La_3Ni_2O_7}$ reveal highly dispersive magnetic excitations whose dispersion is best described by strong interlayer and relatively weak intralayer magnetic couplings \cite{xchen2024}, which is also supported by the inelastic neutron scattering ~\cite{txie2024,hzhou2026}, in sharp contrast to cuprates and many iron-based superconductors where in-plane exchange dominates~\cite{lee2006doping,pdai2015}.

\begin{figure*}
	\centering
	\includegraphics[width=2\columnwidth]{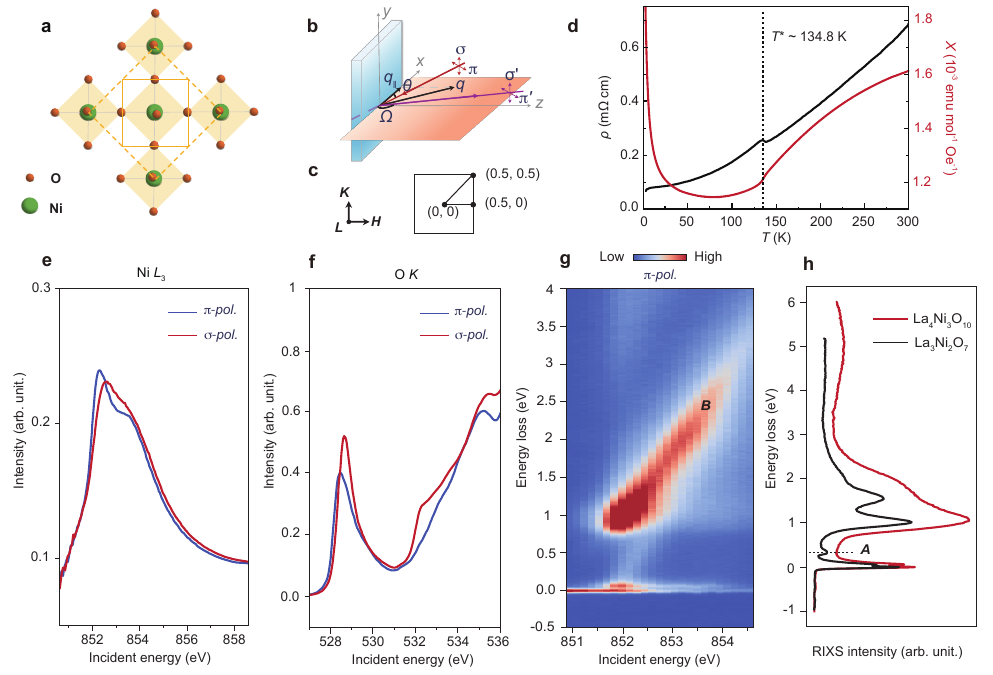}
	\caption{\textbf{Transport and magnetic susceptibility properties, XAS and the incident energy-dependent RIXS map in La$_4$Ni$_3$O$_{10}$.}
		\textbf{a} Top view of the NiO$_2$ plane in La$_4$Ni$_3$O$_{10}$. The solid orange square represents the pseudo-tetragonal unit cell, and the dashed orange square represents the pseudo-orthorhombic $Bmab$ structure (The real space group is $P2_1/a$).
		\textbf{b} Schematic of the RIXS experimental geometry.
        \textbf{c} In-plane Brillouin zone for the pseudo-tetragonal unit cell.
		\textbf{d} Temperature dependence of resistance and magnetic susceptibility at ambient pressure.
		\textbf{e, f} $\pi$ polarized and $\sigma$ polarized XAS spectra of La$_4$Ni$_3$O$_{10}$ for Ni $L_3$-edge (e) and O $K$-edge (f), respectively. Here, the Ni $L_3$-edge XAS has been corrected by subtracting the La $M$-edge contribution below 852 eV.
		\textbf{g}, RIXS intensity map measured as a function of incident photon energy with $\pi$ polarisation.
		\textbf{h} Integrated RIXS spectra in \textbf{g} over the incident energy range of 851.8-853.4~eV (black curve). The corresponding spectra of $\mathrm{La_3Ni_2O_7}$ (red curve) from Ref.~\cite{xchen2024} are overlaid for comparison.}
	\label{fig1}
\end{figure*}

A significant difference in $\mathrm{La_4Ni_3O_{10}}$ comparing to the bilayer compound lies in the number of NiO$_2$ layers. The additional inner NiO$_2$ layer splits the $d_{z^2}$ orbital into bonding, nonbonding and antibonding states through the strong interlayer coupling, and introduces an intrinsic inequivalence between inner and outer Ni sites~\cite{hli2017,jyang2026}. Such trilayer splitting can redistribute the $d_{z^2}$ spectral weight near the Fermi level, impact the hybridization between $3d_{x^2-y^2}$ and $3d_{z^2}$ orbitals which are relevant for the magnetic couplings~\cite{laBollita2024,jli2024,leonov2024,yzhang2024,qyang2024,cchen2024,ptian2024,jwang2024,sakakibara2024b,mzhang2024,clu2025,zhuo2025,gjiang2026}. Recent spectroscopic studies have revealed distinct electronic structure and excitations in $\mathrm{La_4Ni_3O_{10}}$ comparing to $\mathrm{La_3Ni_2O_7}$~\cite{zliu2025,sxu2025,jyang2026,zjiang2026,zhang2025distinct}. Interestingly, the trilayer-induced changes were proposed theoretically as key factors underlying the much lower $T_\mathrm{c}$ of $\mathrm{La_4Ni_3O_{10}}$ than in $\mathrm{La_3Ni_2O_7}$~\cite{ptian2024,jwang2024,mzhang2024,sakakibara2024b,clu2025}. Therefore, directly probing the trilayer-modified electronic interaction especially the full characterization of the magnetic excitation is essential for understanding the interlayer magnetic exchange coupling and its impact on superconductivity.

Here, we employ high-resolution Ni $L_3$-edge RIXS to investigate the electronic and magnetic properties of $\mathrm{La_4Ni_3O_{10}}$ single crystals at ambient pressure. We find that, compared to $\mathrm{La_3Ni_2O_7}$, the Ni $L_3$-edge RIXS spectra of $\mathrm{La_4Ni_3O_{10}}$ exhibit broader $dd$ orbital excitations and a more pronounced fluorescence continuum, indicative of weakened electronic correlation. The low-energy spectra reveal quasi-two-dimensional magnetic excitations with both acoustic and optical branches resolved. Linear-spin-wave analysis demonstrates a substantially reduced effective interlayer exchange $J_z$ compared with $\mathrm{La_3Ni_2O_7}$. Moreover, quasi-elastic RIXS directly resolves a long-range incommensurate SDW which may be linked with its relatively strong itinerancy. Our results underscore the critical role of the electron correlation and the interlayer magnetic coupling for superconductivity in RP nickelates.

\section{Results}
The crystal structure of $\mathrm{La_4Ni_3O_{10}}$ is monoclinic $P2_1/a$ at room temperature, with a competing metastable orthorhombic $Bmab$ phase~\cite{jzhang2020a}. For convenience, reciprocal space indices $(H, K, L)$ are defined in reciprocal lattice units (r.l.u.) based on a pseudo-tetragonal unit cell (Fig.~\ref{fig1}a). The scattering geometry is shown in Fig.~\ref{fig1}b-c, in which the incident X-ray is linearly polarised. Figure~\ref{fig1}d shows the resistivity and magnetic susceptibility of the $\mathrm{La_4Ni_3O_{10}}$ single crystal used in RIXS measurements, both showing a pronounced anomaly at around 135 K, indicating the transition of spin-charge density wave order, consistent with previous reports~\cite{jzhang2020a,yzhu2024}.

\begin{figure*}
	\centering
	\includegraphics[width=2\columnwidth]{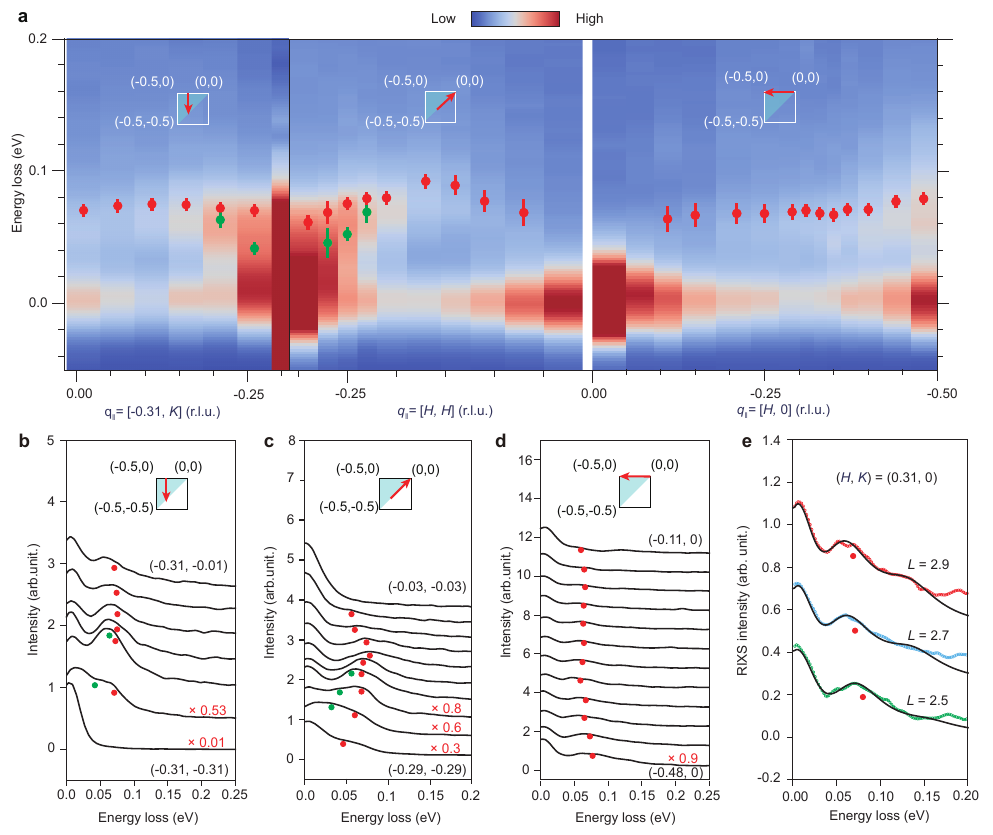}
	\caption{\textbf{The dispersion of magnon in La$_4$Ni$_3$O$_{10}$.}
		\textbf{a} RIXS intensity maps along three high-symmetry directions as indicated with red arrows in the insets. Measurements were taken at 20 K using $\sigma$ polarisation at the Ni $L_3$-edge of 852.3 eV. The red- and green-filled circles denote the undamped energies of the magnetic excitations, extracted from DHO fits, in this and all other panels of the figure. The error bars combine the uncertainty from fitting the RIXS spectra and the experimental energy resolution.
		\textbf{b-d} RIXS spectrum at representative projected in-plane momentum transfers. Intensities of some curves are multiplied by constant factors, as indicated.
		\textbf{e} $L$ dependence of RIXS spectra at the fixed $q_\parallel =$ (-0.31, 0). Black curves show the fitted RIXS spectra, and red filled circles indicate the fitted undamped magnon energies.
	}
	\label{fig2}
\end{figure*}

In Fig.~\ref{fig1}e, f, we present XAS spectra acquired near the Ni $L_3$-edge and O $K$-edge, respectively. The Ni $L_3$ spectra exhibit a sharp resonant peak followed by a higher energy satellite peak, closely resembling that of $\mathrm{La_3Ni_2O_7}$~\cite{xchen2024}. At the O $K$-edge, a significant pre-peak appears under both $\sigma$ and $\pi$ polarisations, indicative of ligand holes on planar O $p_{x,y}$ and apical O $p_z$ states that hybridize with Ni $3d_{x^2-y^2}$ and $3d_{z^2}$, respectively. Figure~\ref{fig1}g displays an incident-energy RIXS map of $\mathrm{La_4Ni_3O_{10}}$ across the Ni $L_3$-edge. When tuning to the vicinity of the Ni $L_3$ resonance of $\sim$~852.3 eV, a prominent excitation at $\sim$~1 eV of energy loss appears, corresponding to the $t_{2g}$ to $e_g$ $dd$ orbital excitation~\cite{xchen2024}. For incident energies above 852.5 eV, a strong fluorescence-like response emerges, attributable to a delocalized electron-hole continuum~\cite{bisogni2016,ylu2018}. Furthermore, a low-energy excitation $\sim$~0.1 eV energy loss is observed, whose intensity peaks at the Ni $L_3$ resonance, which will be discussed in the next section.

\begin{figure*}
	\centering
	\includegraphics[width=2\columnwidth]{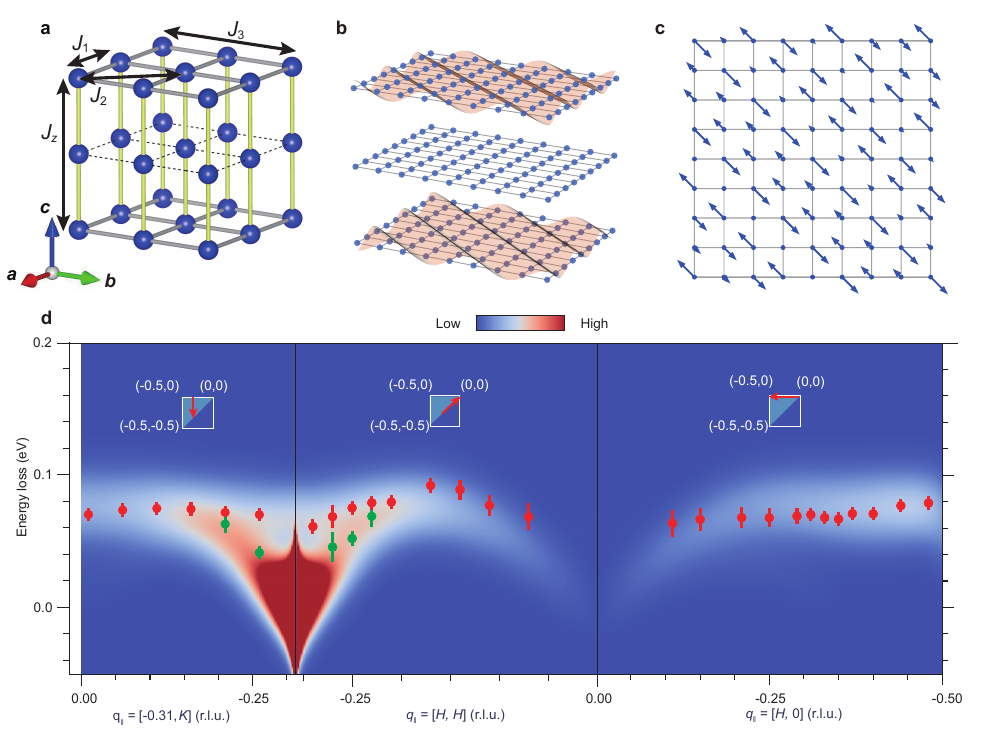}	
	\caption{\textbf{The spin configuration and the calculated dispersion of magnon in La$_4$Ni$_3$O$_{10}$.}
		\textbf{a} Schematic illustration of the in-plane exchange couplings $J_1-J_2-J_3$ and interlayer exchange coupling $J_z$ characterizing the effective Heisenberg Hamiltonian for La$_4$Ni$_3$O$_{10}$. To simplify the sketch only nickel cations are shown.
		\textbf{b} The SDW order of one trilayer as illustrated in the ref. \cite{jzhang2020b} is shown. The SDW is only present in the top and bottom layers.
		\textbf{c} The spin configuration for the SDW order in one outer layer. Arrows represent the magnetic moments. The magnitude of each magnetic moment is reflected by the arrow length.  
		\textbf{d} The experimental magnon dispersion $\epsilon_q$ (red filled circles) versus projected in-plane momentum transfer $q_\parallel$ along high symmetry directions at 20 K. The dispersion of magnon calculated by a $J_1-J_2-J_z$ Heisenberg model based on the SDW model shown in \textbf{b} is overlaid.
	}
	\label{fig3}
\end{figure*}

For a direct comparison with $\mathrm{La_3Ni_2O_7}$, Ni $L_3$-edge RIXS spectra of both compounds are shown in Fig.~\ref{fig1}h. At the fist glance,  $\mathrm{La_4Ni_3O_{10}}$ shows much broader spectral profile comparing to that of $\mathrm{La_3Ni_2O_7}$. Also, the fluorescence excitation (feature B) is more prominent in $\mathrm{La_4Ni_3O_{10}}$  \cite{xchen2024}.  Both phenomena suggest a stronger electron itinerancy in the trilayer PR nickelate than the bilayer counterpart~\cite{bertinshaw2021,rossi2021}. Interestingly, the lowest $dd$ orbital excitation shown in $\mathrm{La_3Ni_2O_7}$ (feature A $\sim$ 0.4 eV ) is absent in $\mathrm{La_4Ni_3O_{10}}$. This excitation involves the transition between the 3$d_{z^2}$ molecular orbital bonding state and the 3$d_{x^2-y^2}$ orbital state, and is highly sensitive to the hopping strength of the interlayer 3$d_{z^2}$ orbitals mediated by apical oxygen $p_z$ states~\cite{xchen2024}. The negligible feature A may result from the trilayer splitting and more itinerancy of $\mathrm{La_4Ni_3O_{10}}$~\cite{tenhuisen2025}. By contrast, the $dd$ excitations between 0.6 eV and  3 eV show nearly the same energy splitting, i.e., 3$d$ crystal field energy of the two materials~\cite{xchen2024}. 

\subsection*{Magnetic excitations}

\begin{figure*}
	\centering
	\includegraphics[width=2\columnwidth]{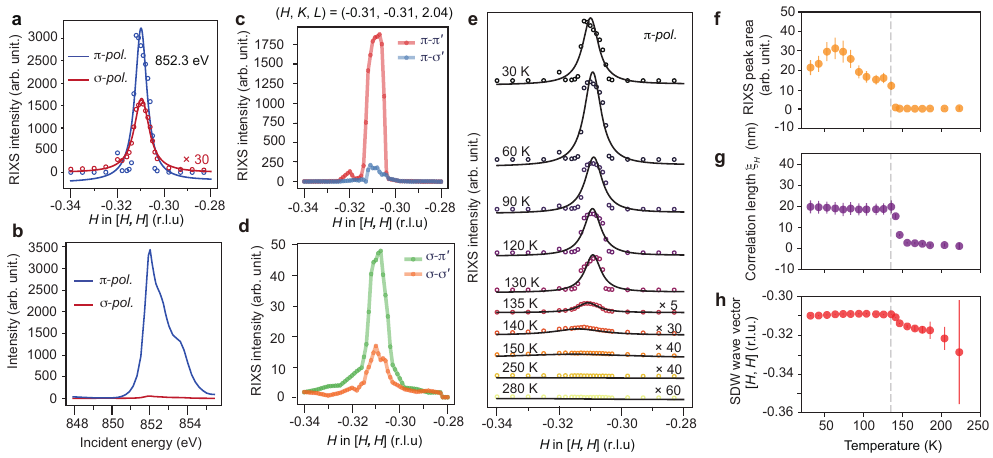}
	\caption{\textbf{SDW order in La$_4$Ni$_3$O$_{10}$.}
		\textbf{a} Integrated SDW peak intensities as a function of projected momentum transfer ($q_\parallel$) along the ($H,H$) direction for $\pi$ and $\sigma$ polarisations. The integration is performed over an energy loss window of $\pm 20$~meV, corresponding to the energy resolution. The solid curves represent Lorentzian fits. $\sigma$-polarized data are scaled for clarity. 
        \textbf{b} SDW peak intensities as a function of incident photon energy and polarization.  
        \textbf{c, d} Polarimetric SDW data. The spectra are decomposed into $\pi\pi'$, $\pi\sigma'$, $\sigma\pi'$, and $\sigma\sigma'$ channels, corresponding to the incoming and outgoing polarization components, respectively. See Methods and Ref.~\cite{xchen2024} for details.  
        \textbf{e} Temperature dependence of the SDW peak. The solid curves represent Lorentzian fits. Intensities were integrated over an energy loss window of $\pm 20$~meV.
        \textbf{f-h} Temperature dependence of the SDW peak area (\textbf{f}), correlation length (\textbf{g}), and SDW wave vector position (\textbf{h}) obtained from the fits in \textbf{e}. Error bars indicate the standard deviation from Lorentzian fits.
	}
	\label{fig4}
\end{figure*}

Figure \ref{fig2} shows the detailed momentum dependence of the RIXS spectra measured at the Ni $L_3$ resonance of 852.3 eV.
The low-energy excitation emanates from the $\Gamma$ point and becomes near-flat along ($H$, 0). Along $(H, H)$, it reaches the maximal energy of $\sim$ 90 meV at (0.17, 0.17) then gets softened towards (0.31, 0.31), where a quasi-elastic scattering peak appears. Previous neutron diffraction measurements on $\mathrm{La_4Ni_3O_{10}}$ have revealed an incommensurate SDW order below the phase transition temperature with a magnetic propagation wavevector at $\boldsymbol{q}_s$ = (0.31, 0.31) \cite{jzhang2020b}. Indeed, the polarisation study on the quasi-elastic scattering peak confirms its magnetic origin (Fig.~\ref{fig4}c,d). The smooth evolution of the low-energy excitations towards the SDW, together with the similar bandwidth to that of $\mathrm{Nd_4Ni_3O_{10}}$ \cite{tenhuisen2025} support the magnetic origin of the low-energy excitations. Note that the magnon has a smaller bandwidth than that of the single-layer $\mathrm{La_2NiO_4}$ (120 meV ) \cite{petsch2023}. The magnon exhibits negligible dispersion along the out-of-plane direction, indicating a quasi-two-dimensional nature (Fig. \ref{fig2}e).

To determine the dispersion of the magnetic excitations, we extracted the undamped energy of the magnon by fitting the spectra with a damped harmonic oscillator (See details in Supplementary Note 2). Both acoustic and optical branches of magnetic excitations are clearly resolved in the vicinity of $\boldsymbol{q}_s$, as seen more clearly in spectra taken at momentum near $\boldsymbol{q}_s$ (Fig. \ref{fig2}b-c). The acoustic branch softens approximately linearly and extrapolates to zero energy at $\boldsymbol{q}_s$, with the optical branch appearing at high energy. This behavior contrasts sharply with $\mathrm{La_3Ni_2O_7}$, where only the acoustic branch is observed. Near the $\Gamma$ point, both branches become indiscernible because of significantly reduced spectral weight.

In $\mathrm{La_4Ni_3O_{10}}$, a model of the incommensurate SDW order was proposed in Ref. \cite{jzhang2020b} using neutron diffraction, as illustrated in Fig. \ref{fig3}b,c. In the trilayer structure, the central NiO$_2$ plane lies at an SDW node, i.e., carries no static ordered moment, whereas the two outer planes host antiferromagnetically aligned order moments with respect to each other \cite{jzhang2020b,mzhang2025a}. Within each outer layer, the in-plane spin modulation can be written as $\cos(\boldsymbol{q}_s\cdot\boldsymbol{r}_{\perp})$, where $\boldsymbol{r}_{\perp}$ denotes the in-plane coordinates of the Ni sites. The ordered moments are confined to the NiO$_2$ planes and are oriented perpendicular to $\boldsymbol{q}_s$ \cite{jzhang2020b}.

Although $\mathrm{La_4Ni_3O_{10}}$ has appreciable itinerant character, we fit the extracted magnon dispersion to obtain phenomenologically the effective dominant magnetic interactions using the linear-spin-wave theory (LSWT) in the SpinW package~\cite{toth2015}. We assume an effective interlayer exchange interaction $J_z$ that couples the top and bottom layers. The Heisenberg model is written as
\begin{equation}
	H = J_z \sum_i \mathbf{S}_{i,t}\!\cdot\!\mathbf{S}_{i,b}
	+ \sum_{n=1}^{3} J_n \sum_{\langle ij\rangle_n,\alpha} \mathbf{S}_{i,\alpha}\!\cdot\!\mathbf{S}_{j,\alpha},
\end{equation}
where $i$, $j$ label in-plane sites, $\alpha$ is the layer index for the top (t) and bottom (b) layers, and $\langle ij\rangle_n$ denotes $n$th-neighbor pairs within a layer. $J_1$, $J_2$ and $J_3$ are the in-plane nearest-, next-nearest-, and third-neighbor exchange couplings, respectively (Fig. \ref{fig3}a). In a bilayer model coupled by $J_c$, the magnetic excitation intensity of the acoustic and optical branches are modulated with the out-of-plane component $L$ as $I_{\mathrm{ac}}(L) \varpropto$ sin$^2(z\pi L)$ and $I_{\mathrm{op}}(L) \varpropto$ cos$^2(z\pi L)$, respectively, where $zc$ is the intra-bilayer spacing and $c$ is the lattice parameter. This $L$ modulation of the acoustic and optical branches is also a generic feature of bilayer cuprates, where odd- and even-parity magnetic excitations have been widely observed, such as $\mathrm{YBa_2Cu_3O_{6+\delta}}$ and $\mathrm{Bi_2Sr_2CaCu_2O_{8+\delta}}$~\cite{keimer1998bilayer,capogna2007}. In our RIXS measurements, the scattering angle $2\theta$ between the incident and scattered X-rays was fixed (Fig. \ref{fig1}b). Consequently, scanning the in-plane projected momentum transfer by varying the incident angle $\theta_i$ necessarily produces a concomitant change of the out-of-plane component $L$. To enable quantitative comparisons of the theory with the experiment, the calculated intensities explicitly include this $L$-dependent modulation. In addition, for an incommensurate magnetic structure with a single propagation vector $\boldsymbol{q}$, the dynamical structure factor may contain not only the main branch at $\boldsymbol{Q}$, but also additional replica-like branches at $\boldsymbol{Q}\pm\boldsymbol{q}$ arising from the incommensurate modulation \cite{toth2015,janoschek2010}. Accordingly, the LSWT calculation includes six dispersive branches in the simulated spectrum.

The simulated dynamic structure factor $S_{xx}+S_{yy}+S_{zz}$, calculated using the optimized  exchange coupling $SJ_1 = 15.7 \pm 1.2$ meV, $SJ_2 = -0.3 \pm 0.2$~meV, $SJ_3 = 10.8 \pm 0.8$~meV, and $SJ_z = 22.3 \pm 5.5$~meV, is shown in Fig. \ref{fig3}d. To facilitate comparison with the experiment, the spectrum is convoluted with the instrumental energy resolution. The resulting spectrum reproduces both the dispersion and the spectrum-weight distribution observed experimentally, indicating overall a good agreement with the measured data. The fitted $SJ_1$ is smaller than that of $\mathrm{La_2NiO_4}$~\cite{petsch2023}, consistent with the reduced magnon bandwidth in $\mathrm{La_4Ni_3O_{10}}$. The most salient difference relative to $\mathrm{La_3Ni_2O_7}$ is the significantly reduced interlayer magnetic interaction $SJ_z$ and enhanced in-plane interactions.

\subsection*{Spin-density-wave order}

We performed incident X-ray polarisation analysis of the quasi-elastic scattering near $\boldsymbol{q}_s$. A pronounced peak is observed at $\boldsymbol{q}_s$ with incident photon energy tuned to the Ni $L_3$ resonance of $\sim$~852.3 eV as shown in Fig.~\ref{fig4}a and b, providing direct evidence for static SDW order. Along the [$H$, $H$] direction, a Lorentzian fit to the $\pi$-polarized data yields a full width at half maximum (FWHM) of 0.00623~$\pm$~0.00097 r.l.u, corresponding to an in-plane correlation length $\xi_H = 2/\mathrm{FWHM} \approx$ 20~nm. This is about 51 times the Ni-O-Ni bond length, indicating long-range SDW order within our experimental resolution. The incident-energy dependence further shows that the SDW peak resonates strongly at the Ni $L_3$-edge but remains negligible at the La $M_4$ edge, despite the La $M_4$ edge being stronger in the XAS. This mismatch illustrates that the SDW scattering predominantly originates from Ni~3$d$~-~O~2$p$ hybridized states rather than La-derived states. The SDW intensity is also highly polarisation selective ($I_\pi/I_\sigma$ $\approx$ 62), suggesting $d_{z^2}$ orbital dominance. By analyzing the polarisation of incoming and outgoing X-rays, the scattering intensity under $\pi$-$\pi'$, $\pi$-$\sigma'$, $\sigma$-$\pi'$ channels is substantially stronger than the $\sigma$-$\sigma'$ channel (Fig.~\ref{fig4}c,d), confirming the magnetic origin \cite{xchen2024}. 

The temperature dependence of the SDW peak is summarized in Fig.~\ref{fig4}e-h. With increasing temperature, both the correlation length and the SDW wavevector remain essentially constant up to $\sim$~135~K. Above this temperature, the peak intensity and the correlation length $\xi_H$ abruptly decay to nearly zero, while the SDW wavevector shifts away from $\boldsymbol{q}_s$, suggesting the vanishing of the long-range SDW order, which is consistent with our transport and magnetic susceptibility results, as well as previous neutron scattering studies on $\mathrm{La_4Ni_3O_{10}}$ \cite{jzhang2020b}. In addition to $\sim$~135~K, we revealed a second characteristic temperature around 60 K, at which the peak intensity reaches its maximum. A similar behavior was observed in $\mathrm{La_3Ni_2O_{7-\delta}}$ thin film in the $\sigma$ channel, where it was ascribed to the charge anisotropy~\cite{xren2025}. This suggests that, in $\mathrm{La_4Ni_3O_{10}}$, the interplay and possible competition between intertwined SDW and charge degree of freedom may be a source of the anomaly at $\sim$~60~K.

\section*{Discussion}

By directly comparing $\mathrm{La_4Ni_3O_{10}}$ with $\mathrm{La_3Ni_2O_7}$, our RIXS measurements demonstrate that the trilayer nickelate exhibits distinct electronic and magnetic properties. Our results establish that $\mathrm{La_4Ni_3O_{10}}$ is more itinerant and hosts a substantially reduced interlayer exchange coupling. The itinerancy is reinforced by the broadening of the $dd$ orbital transitions and more pronounced fluorescence-like excitations~\cite{bertinshaw2021,rossi2021}. Previous optical spectroscopy studies reported that the ratio of experimental to theoretical kinetic energy in $\mathrm{La_4Ni_3O_{10}}$ is about an order of magnitude larger than in $\mathrm{La_3Ni_2O_7}$, placing $\mathrm{La_4Ni_3O_{10}}$ in the regime of a moderately correlated metal, which is consistent with our RIXS results~\cite{zliu2025,sxu2025}. Theoretical calculations suggested layer-dependent electron correlation~\cite{jwang2024,zhuo2025}: the inner NiO$_2$ layer is more hole-doped and weakly correlated, corresponding to a higher Ni valence, whereas the outer NiO$_2$ layer remains more strongly correlated, although slightly less so than in $\mathrm{La_3Ni_2O_7}$ due to their increased hole doping. Our RIXS measurements do not resolve individual layers, but the enhanced itinerancy in $\mathrm{La_4Ni_3O_{10}}$ is naturally explained by this layer-dependent scenario, with the weakly correlated inner layer promoting itinerant behavior.

Our quasi-elastic RIXS data directly resolve long-range SDW order at $\boldsymbol{q}_s$ = (0.31, 0.31), which disappears near 135 K, consistent with previous neutron diffraction results~\cite{jzhang2020b}.  Combined with recent single-crystal $\mu$SR~\cite{ycao2025}, our data further support that the density-wave state in $\mathrm{La_4Ni_3O_{10}}$ is an amplitude-modulated SDW with a spatially varying spin density, more reminiscent of chromium. This is notably different from bilayer $\mathrm{La_3Ni_2O_7}$, where the consensus is formed such that a bicollinear stripe order with fixed local moments~\cite{gupta2025}. More recently, ARPES resolved the momentum-dependent SDW gap, and combined with tight-binding and functional renormalization group analysis, revealed a mirror-selective nesting mechanism in which opposite-parity bands generate an interlayer-antiferromagnetic SDW instability at $\boldsymbol{q}_s$ between the outer NiO$_2$ layers~\cite{jyang2026,zjiang2026}. This amplitude-modulated nature of the SDW, together with the nesting driven origin, is consistent with the more itinerant electronic character of $\mathrm{La_4Ni_3O_{10}}$.

In unconventional superconductors, the on-site Coulomb repulsion, a measure of the electronic correlation, is one of the key parameters setting the relevant energy scale of superconductivity. For instance, the parent compound of cuprates, with the highest ambient-pressure $T_c$, is an antiferromagnetic Mott insulator. In the iron-based superconductors, which has a lower maximal ambient-pressure $T_c$, the parent compound features spin density wave and is a bad metal. The same trend can be seen in more recent examples of Kagome superconductors: CsCr$_3$Sb$_5$ has a higher $T_c$ than its counterpart CsV$_3$Sb$_5$ in which the former is characterized with relatively stronger electronic correlation than the latter.  \cite{liu2024superconductivity}.

The central result of our RIXS spectra is the observation of quasi-two-dimensional magnetic excitations containing sizable acoustic and optical modes. This contrasts with $\mathrm{La_3Ni_2O_7}$, where the magnon is dominated by the acoustic branch~\cite{xchen2024}. For $\mathrm{La_4Ni_3O_{10}}$, linear-spin-wave analysis yields effective exchange interactions of $SJ_1 = 15.7 \pm 1.2$ meV, $SJ_2 = -0.3 \pm 0.2$~meV, $SJ_3 = 10.8 \pm 0.8$~meV, and $SJ_z = 22.3 \pm 5.5$~meV, indicating pronounced in-plane magnetic coupling and a significantly reduced interlayer exchange. The $J_z$ is only about 30$\%$ of that in $\mathrm{La_3Ni_2O_7}$. Interestingly, the ratio of $T_\mathrm{c}$ between the trilayer and bilayer nickelates is comparable to that of $J_z$. Recently, rare-earth substitution in La$_{3-x}$(Nd, Sm)$_x$Ni$_2$O$_7$ shows the enhancement of $T_\mathrm{c}$ to above 90~K by compressing the $c$-axis lattice parameter, which is expected to strengthen the interlayer magnetic exchange~\cite{zqiu2025,qzhong2025}. These results suggest the importance of interlayer exchange for superconductivity in RP nickelates. By contrast, in the trilayer square-planar compounds $R_4\mathrm{Ni_3O_8}$, the dominant magnetic interaction is the in-plane exchange $J_1$ owing to the absence of apical oxygen, even though they exhibit a magnon energy scale comparable to that of $\mathrm{La_4Ni_3O_{10}}$~\cite{jlin2021}. Notably, superconductivity has not been established in $R_4\mathrm{Ni_3O_8}$ under either high-pressure condition~\cite{jcheng2012} or in thin-film form~\cite{segedin2023}. This comparison further underscores the role of interlayer magnetic coupling in superconductivity in RP nickelates. The mechanism appears to be distinct from that in the cuprate family. In cuprates, the $d_{x^2-y^2}$ orbital plays a crucial role in superconducting pairing, and the highest $T_\mathrm{c}$ is realized in trilayer compounds, where layer differentiation creates an advantageous combination of underdoped inner planes and overdoped outer planes, enhancing pairing strength and phase stiffness simultaneously~\cite{xluo2023}.

~\\
\noindent{\bf Methods}

\noindent{\bf Sample preparation}

The $\mathrm{La_4Ni_3O_{10}}$ sample was synthesized using the high-oxygen-pressure floating-zone technique, as described in Ref.~\cite{jli2024}. The sample quality was characterized by X-ray diffraction (XRD) and Laue diffraction (see Supplementary Information for details). Prior to the RIXS measurements, the sample was cleaved in order to obtain a flat and clean surface.

\vspace{3mm}
\noindent{\bf XAS and RIXS measurements}

X-ray absorption spectroscopy (XAS) and resonant inelastic X-ray scattering (RIXS) measurements were conducted at the I21 beamline of Diamond Light Source \cite{zhou2022i21}. The crystal structure of $\mathrm{La_4Ni_3O_{10}}$ is described in a pseudo-tetragonal setting with lattice parameters $a_{\rm T}$ = $b_{\rm T}$ $\approx 3.87\,\mathrm{\AA}$ and $c = 28.5\,\mathrm{\AA}$, defining reciprocal space in units where $2\pi/a_{\rm T} = 2\pi/b_{\rm T} = 2\pi/c = 1$. Momentum transfer is expressed as $\mathbf{Q} = (H, K, L)$ in reciprocal lattice units.

The sample was mounted such that the $a_T$-$c$ plane lies in the horizontal scattering geometry, with crystallographic alignment established via Bragg and ordering reflections to ensure that the ${c}^*$ axis resides within the scattering plane. Unless otherwise specified, the spectrometer arm was fixed at a scattering angle of $\Omega = 154^\circ$.

XAS spectra were acquired at 20 K in total electron yield mode under grazing incidence conditions ($\theta_0 = 20^\circ$), enabling sensitivity to both in-plane and out-of-plane electronic states. The measurements were normalized to the incident photon flux, and both $\sigma$ and $\pi$ linear polarisations were employed.

For RIXS, energy-dependent measurements were performed across the Ni $L_3$-edge (850.9-854.5 eV) to map out the full resonant response, with an overall energy resolution of 39.5 meV (FWHM). The scattering geometry was chosen in the grazing-in configuration to optimize sensitivity to low-energy excitations, and both polarisation channels were measured.

polarisation analysis of the scattered photons was carried out using a graded multilayer analyzer optimized for the Ni $L_3$ edge. Measurements at selected momentum transfers $\lvert q_s \rvert$ = $(0.31, 0.31, L)$ were used to investigate excitations associated with the spin density wave (SDW) ordering vector. The overall energy resolution in this configuration was approximately 54 meV (FWHM). Due to the deviation from the ideal Brewster condition, the reflected signal represents a mixture of different linear polarisation components. Accordingly, the measured direct and indirect RIXS intensities can be expressed following the formulation given in Ref.~\cite{xchen2024}.

\vspace{3mm}
\noindent{\bf LSWT analysis}

The magnetic excitation spectra were analyzed using linear spin-wave theory for a bilayer Heisenberg model. Numerical calculations of the dynamical structure factor were performed with SpinW, and the exchange parameters were obtained by least-squares fitting to the experimental dispersion. Full details of the model, fitting procedure, and analytical LSWT expressions are provided in the Supplementary Note 3.

\vspace{5mm}
\noindent{\bf Data availability}

All data generated or analyzed during this study are available from the corresponding authors upon reasonable request.

%

\vspace{5mm}
\noindent{\bf Acknowledgments}

We thank Xianhui Chen, G. Sawatzky, and Kun Jiang for fruitful discussions. We acknowledge Diamond Light Source for providing beamtime at I21 Beamline under Proposal NR42450. D.L.F. acknowledges the fundings from Quantum Science and Technology-National Science and Technology Major Project (Grant No.: 2021ZD0302803) and the New Cornerstone Science Foundation, China (Grant No. NCI202211). Y.L. acknowledges support from the National Natural Science Foundation of China (No. 12274207), the National Key R\&D Program of China (No. 2022YFA1403000), and the Basic Research Program of Jiangsu (No. BK20253009). Work at SYSU was supported by the National Natural Science Foundation of China (Grant No. 12425404, U25A20193), the Fundamental and Interdisciplinary Disciplines Breakthrough Plan of the Ministry of Education of China (JYB2025XDXM403), the Guangdong Major Project of Basic Research (2025B0303000004), the Guangdong Provincial Key Laboratory of Magnetoelectric Physics and Devices (Grant No. 2022B1212010008), and Research Center for Magnetoelectric Physics of Guangdong Province (Grant No. 2024B0303390001). Y-.F.C. acknowledges funding from Diamond Light Source and the Clarendon Scholarship from the University of Oxford under joint doctoral studentship No. STU0477.

\vspace{5mm}
\noindent{\bf Author contributions}

K.-J.Z. and D.L.F. initiated the project. X.Y.C. and Y.-F.C. performed the Laue measurement. X.Y.C., M.X., M.G.F., S.A., W.L.Z. and K.-J.Z. conducted XAS and RIXS experiments at Diamond Light Source. X.Y.C., M.X., Z.Z.L., and K.-J.Z. analysed and interpreted the data. Z.Z.L. and Y.L. performed the LSWT calculations. D.Y.H. and M.W. grew the sample and performed the transport and magnetic susceptibility measurements. Z.Z.L., K.-J.Z., M.X. and X.Y.C. wrote the manuscript with input from all authors. K.-J.Z. is responsible for project direction and planning.

\vspace{5mm}
\noindent{\bf Competing interests}

The authors declare no competing interests.

\end{document}